\documentclass[11pt,a4paper]{article}

\usepackage[]{hyperref}
\usepackage{amsmath, amsthm, amssymb}
\usepackage{graphicx}

\usepackage{setspace}

\def\ket#1{\left|#1\right>}
\def\bra#1{\langle#1\vert}
\def\exptt#1{\langle#1\rangle}
\def\expt(#1#2){\langle #2 \vert #1 \vert #2 \rangle}

\title{A no-go result for QBism} 
\author{Shan Gao
\\Research Center for Philosophy of Science and Technology, 
\\ Shanxi University, Taiyuan 030006, P. R. China
\\ E-mail:  \href{mailto:gaoshan2017@sxu.edu.cn}{gaoshan2017@sxu.edu.cn}.}

\begin{document}
% \begin{spacing}{2.0}
\maketitle
%\vspace{2mm}
%\newpage
%\tableofcontents

\begin{abstract}\noindent 

In QBism the wave function does not represent an element of physical reality external
to the agent, but represent an agent's personal probability assignments, reflecting his subjective degrees of belief about the future content of his experience. In this paper, I argue that this view of the wave function is not consistent with protective measurements. The argument does not rely on the realist assumption of the $\psi$-ontology theorems, namely the existence of the underlying ontic state of a quantum system. 

\end{abstract}

\vspace{6mm}

QBism is a new interesting approach to understanding quantum mechanics (Caves, Fuchs and Schack, 2002, 2007; Fuchs, Mermin and Schack, 2013; Fuchs and Stacey, 2016). It has received more and more attention in recent years (see, e.g. Mermin, 2014).  
In QBism the wave function represents an agent's personal probability assignments, reflecting his subjective degrees of belief about the future content of his experience such as the outcome of a measurement. 
Since QBism admits no agent-independent elements of physical reality that determine either measurement outcomes or probabilities of measurement outcomes, it is still consistent with the $\psi$-ontology theorems, which can be proved only when assuming the existence of the underlying ontic state of a quantum system (Pusey, Barrett and Rudolph, 2012; Colbeck and Renner, 2012; Hardy, 2013). 
However, it has been debated whether QBism provides a fully successful new framework for understanding quantum mechanics (Timpson, 2008; Bacciagaluppi 2014; Norsen, 2016; McQueen, 2017; Earman, 2019). 
In this paper, I will argue that the interpretation of the wave function according to QBism is not consistent with protective measurements. The argument does not rely on the realist assumption of the existence of the underlying ontic state of a quantum system. 

Protective measurement (PM) is a method to measure the expectation value of an observable on a single quantum system (Aharonov and Vaidman, 1993; Aharonov, Anandan and Vaidman, 1993; Vaidman, 2009; Gao, 2015; Piacentini et al, 2017). 
For a conventional projective measurement, the measurement outcome will be in general random, being an eigenvalue of the measured observable with a probability in accordance with the Born rule, and the expectation value of the observable can be obtained only as the statistical average of eigenvalues for an ensemble of identically prepared systems. 
By contrast, during a PM the wave function of  the measured system is protected by an appropriate procedure so that it keeps unchanged during the measurement.\footnote{There are two known schemes of protection. The first one is to introduce a protective potential, and the second one is via the quantum Zeno effect. It should be pointed out that the protection requires that some information about the measured system should be known before a PM, and PMs cannot measure an arbitrary unknown wave function. In some cases, the information may be very little. For example, we only need to know that a quantum system such as an electron is in the ground state of an external potential before we make PMs on the system to find its wave function, no matter what form the external potential has.} 
Then, by the deterministic Schr\"{o}dinger evolution that is independent of the Born rule, the measurement outcome will be definite, being the expectation value of the measured observable, even if the system is initially not in an eigenstate of the observable. 

This result can be seen from the following simple derivation. 
As for a projective measurement, the interaction Hamiltonian for measuring an observable $A$ is given by the usual form $H_I = g(t)PA$, where $g(t)$ is the time-dependent coupling strength of the interaction, which is a smooth function normalized to $\int_0^T g(t)dt =1$ during the measurement interval $T$, and $g(0)=g(T)=0$, and $P$ is the conjugate momentum of the pointer variable $X$. 
When the wave function of the measured system is protected to keep unchanged during the measurement, the evolution of the wave function of the combined system is

\begin{equation}
\ket{\psi(0)}\ket{\phi(0)} \rightarrow \ket{\psi(t)}\ket{\phi(t)}, t>0,
\end{equation}

\noindent where $\ket{\phi(0)}$ and $\ket{\phi(t)}$ are the wave functions of the measuring device at instants $0$ and $t$, respectively, $\ket{\psi(0)}$ and $\ket{\psi(t)}$ are the wave functions of the measured system at instants $0$ and $t$, respectively, and $\ket{\psi(t)}$ is the same as $\ket{\psi(0)}$ up to an overall phase during the measurement interval $[0,T]$. Then we have

\begin{eqnarray}
{d \over dt}\bra{\psi(t)\phi(t)}X\ket{\psi(t)\phi(t)} &=& {1\over {i\hbar}}\bra{\psi(t)\phi(t)}[X, H_I]\ket{\psi(t)\phi(t)}\nonumber\\
&=&g(t)\bra{\psi(0)}A\ket{\psi(0)},
\end{eqnarray}

\noindent This further leads to

\begin{equation}
\bra{\phi(T)}X\ket{\phi(T)}- \bra{\phi(0)}X\ket{\phi(0)}= \bra{\psi(0)}A\ket{\psi(0)},
\end{equation}

\noindent which means that the shift of the center of the pointer wave packet is the expectation value of $A$ in the initial wave function of the measured system.  
This clearly demonstrates that the result of a measurement of an observable on a  system, which does not change the wave function of the system, is the expectation value of the measured observable in the wave function of the measured system.\footnote{Note that PMs are different from quantum non-demolition measurements. In a quantum non-demolition measurement, the measured observable is required to commute with the total Hamiltonian so that it is a constant of the motion. This implies that the measurement is repeatable, but it does not imply that the wave function of the measured system is unchanged during the measurement. By comparison, a PM does not require that the measured observable must commute with the total Hamiltonian, and the wave function of the measured system does not change during the measurement.} 

Since the wave function can be reconstructed from the expectation values of a sufficient number of observables, the wave function of a single quantum system can be measured by a series of PMs in principle (see Aharonov, Anandan and Vaidman, 1993 for a more detailed analysis). Let the explicit form of the measured wave function at a given instant $t$ be $\psi(x)$, and the measured observable $A$ be (normalized) projection operators on small spatial regions $V_n$ having volume $v_n$:

\begin{equation}
A= 
\begin{cases} 
{1\over{v_n}},& \text{if $x \in V_n$,}
\\
0,&\text{if $x \not\in V_n$.} 
\end{cases}
\label{OA}
\end{equation}

\noindent  A PM of $A$ then yields 

\begin{equation}
\exptt{A} = {1\over {v_n}} \int_{V_n}|\psi(x)|^2 dv,
\end{equation}

\noindent  which is the average of the density $\rho(x) = |\psi(x)|^2$ over the small region $V_n$. Similarly, we can measure another observable $B ={\hbar \over{2mi}} (A\nabla + \nabla A)$, which is the current density at $x$.\footnote{ An example of how to measure $B$ is given by Aharonov, Anandan and Vaidman (1993, p.4622). In the gedanken experiment, a charged particle $Q$ is in a thin circular tube enclosing a magnetic but with the magnetic field vanishing inside the tube. A protective measurement of each eigenstate can be made by shooting electrons near the tube and observing their trajectories; from the accelerations of the electrons the charge density $Q\rho$ and the current density $Qj$ can be determined.} The measurement yields

\begin{equation}
\exptt{B} ={1\over {v_n}} \int_{V_n}{\hbar \over{2mi}}(\psi^* \nabla \psi - \psi  \nabla \psi^* )dv = {1\over {v_n}} \int_{V_n}j(x)dv.
\end{equation}

\noindent This is the average value of the flux density $j(x)$ in the region $V_n$. Then when $v_n \rightarrow 0$ and after performing measurements in sufficiently many regions $V_n$ we can measure $\rho(x)$ and $j(x)$ everywhere in space.\footnote{In most cases the measured wave function can be reconstructed only in principle. For a spatial wave function like $\psi(x)$, since one needs to measure the observables A and B in infinitely many points in space, this is an impossible task in practice.} 
Since the wave function $\psi(x)$ can be uniquely expressed by $\rho(x)$ and $j(x)$ (except for an overall phase factor), the whole wave function of the measured system at a given instant can be measured by PMs.% \footnote{There are two known schemes of PM. The first scheme is to introduce a protective potential such that the wave function of the measured system at a given instant, $\ket{\psi}$, is a nondegenerate energy eigenstate of the total Hamiltonian of the system with finite gap to neighboring energy eigenstates. By this scheme, the measurement of an observable is required to be weak and adiabatic. The second scheme is via the quantum Zeno effect. The Zeno effect is realized by making frequent projective measurements of an observable, of which the wave function of the measured system at a given instant, $\ket{\psi}$, is a nondegenerate eigenstate. By this scheme, the measurement of the measured observable is not necessarily weak but weaker than the Zeno projective measurements.}

Now let's analyze possible implications of PMs for the meaning of the wave function and QBism. 
Proponents of PMs argue that since PMs can measure the expectation values of observables and even the wave function on a single quantum system, they provide strong supports for the reality of the wave function or $\psi$-ontology, while some others disagree, and $\psi$-epistemic models have also been proposed to account for PMs (Combes et al, 2018). Recently I showed that although these $\psi$-epistemic models can explain the appearance of expectation values of observables in a single measurement, their predictions are different from those of quantum mechanics for some PMs. Moreover, I gave a proof of the reality of the wave function in terms of PMs under an auxiliary finiteness assumption about the dynamics of the ontic state (Gao, 2020). However, the new proof is still based on the realist assumption of the $\psi$-ontology theorems.  Thus it has no implications for QBism which denies this assumption. 

In the following, I will present a new analysis of QBism in terms of PMs. 
In particular,  I will argue that QBism is inconsistent with PMs, and the argument does not rely on the realist assumption of the $\psi$-ontology theorems. 
First, since the outcome of a PM of an observable on a quantum system being in a superposition of different eigenstates of the observable is always definite, the superposed wave function does not represent probability assignments for PMs, no matter whether  these probabilities are objective or subjective. 
Next, the wave function can be measured by a series of PMs on a single quantum system. This means that the wave function is a representation of the objective outcome of the interaction between the measured system and the measuring device during the PMs, such as the shift of the pointer of a measuring device which makes the PMs.\footnote{But this does not mean that the wave function must be a direct representation of the ontic state of the measured system even if the ontic state exists (Combes et al, 2018). More work still needs to be done to prove this stronger result (Gao, 2020).}  
For example, the modulus squared of the wave function in position $x$, $|\psi(x)|^2$, is a representation of the outcome of the interaction between the measured system and the measuring device during a PM of the projection operator in $x$. 
Since the outcome of a PM being the modulus squared of the wave function is objective and definite, the wave function cannot be subjective degrees of belief of an agent, which may be different for different agents. 

We can also reach the same conclusion by a somewhat different argument. 
According to QBism, the wave function represents subjective degrees of belief of an agent, and thus it is a property of an agent, not of the external world. 
Then, if the wave function can be measured, it can be measured only from the agent, not from the external world. %\footnote{Here there is an interesting idea for QBism. If the wave function is a property of an agent, then it should be in principle measurable by measuring the brain state of the agent. This also means that the wave function in QBism is a property of a single system, not of an ensemble.}
But PMs show that the wave function can be measured by a certain interaction between a quantum system and a measuring device, which are independent of any agent. 
Thus QBism is not consistent with PMs. 
Note that the inconsistency is not only a logical contradiction but also an observable contradiction, since PMs can be realized in experiments. 

To sum up, for PMs, the wave function represents the objective definite outcomes of the interactions between the measured system and the measuring device, and it does not represent probability assignments, either objective or subjective. 
This is against QBism, according to which the wave function represents an agent's personal probability assignments, reflecting his subjective degrees of belief about the outcome of a measurement. 
Since the above argument does not assume that there are agent-independent
elements of physical reality that determine either measurement outcomes or probabilities of measurement outcomes, it is stronger than the $\psi$-ontology theorems in some sense, which can be proved only under this realist assumption. 

Here is a simple explanation for the inconsistencies between QBism and PMs. 
As pointed out by Gao (2017), there are in fact two connections between the mathematical formalism of quantum mechanics and experience. The first connection is the well-known Born rule, and the connection is in general probabilistic. 
The second connection is PMs, and the connection is definite, determined only by the deterministic Schr\"{o}dinger equation and independently of the Born rule. 
Then, the inconsistency between QBism and PM may be understandable. QBism refers only to the first connection, not to the second connection; it aims to interpret the wave function in the Born rule, but it ignores the wave function in PMs. 

As noted before, some information about the measured state needs to be known before a PM can be made. This additional information is not mentioned in the above analysis. Does this information influence the degree of belief of the agent? The answer depends. In general, for a PM, the agent only needs to know that the measured wave function has been protected, while this information does not influence the degree of belief of the agent. In the special case that the agent knows what the measured wave function is, the information will influence and determine the degree of belief of the agent. In either case, QBism still regards the wave function as a representation of an agent's personal probability assignments, and thus the inconsistencies between QBism and PMs still exist. 

Finally, it is worth noting that the above no-go result is also valid for other pragmatist approaches to quantum theory which deny that the theory offers a description or representation of the physical world (Healey, 2017). 
One may think that this result can be avoided by insisting that the measurement outcome is \emph{only} the particular experience of an agent. 
This is a radical solution for many of us, since it will arguably lead to idealism or even solipsism, and it is committed to ``a challenging program of re-building the manifest image of the external world from extremely thin resources (erocentric experiences)'' (Timpson, 2021). 
However, it seems that even this idealistic version of QBism cannot totally avoid the inconsistencies between QBism and PMs, since according to PMs the wave function does not (only) represent probability assignments, while according to QBism it does. 
It remains to be seen if QBists can find some other ways to resolve the inconsistencies. 
% I will investigate this possibility in more detail in future work. It will be interesting to see if QBists can fulfill this great program. 

\section*{Acknowledgments}
I am grateful to the editors and reviewers of this journal for their useful comments and  suggestions. This work is supported by the National Social Science Foundation of China (Grant No. 16BZX021).
 
\section*{References}
\renewcommand{\theenumi}{\arabic{enumi}}
\renewcommand{\labelenumi}{[\theenumi]}
\begin{enumerate}

\item{} Aharonov, Y., Anandan, J. and Vaidman, L. (1993). Meaning of the wave function. Phys. Rev. A 47, 4616. 

\item{} Aharonov, Y. and Vaidman, L.   (1993). Measurement of the Schr\"{o}dinger wave of a single particle, Physics Letters A 178, 38.

\item{} Bacciagaluppi, G. (2014). A critic looks at QBism. In Galavotti, M. C., Dieks, D., Gonzalez, W. J., Hartmann, S., Uebel, T., andWeber, M., editors, New Directions in the Philosophy of Science, pages 403-416. Springer, Cham.

\item{} Caves, C. M., Fuchs, C. A., and Schack, R. (2002). Quantum probabilities as Bayesian probabilities. Physical Review A 65, 022305.
\item{}  Caves, C. M., Fuchs, C. A., and Schack, R. (2007). Subjective probability and quantum certainty. Studies in History and Philosophy of Modern Physics, 38, 255.

\item{} Colbeck, R., and Renner, R. (2012). Is a system's wave function in one-to-one correspondence with its elements of reality? Physical Review Letters, 108, 150402.

\item{} Combes, J., Ferrie, C., Leifer, M. and Pusey, M. (2018). Why Protective Measurement Does Not Establish the Reality of the Quantum State, Quantum Studies: Mathematics and Foundations 5, 189-211.

\item{} Earman, J. (2019). Quantum Bayesianism assessed. The Monist, 102, 403-423.

\item{}  Fuchs, C. A., Mermin, N. D. and Schack, R. (2013). An Introduction to QBism with an Application to the Locality of Quantum Mechanics. arXiv:1311.5253. Am. J. Phys. 82, 749-754. 
\item{} Fuchs, C. A. and Stacey, B. C. (2016). QBism: Quantum Theory as a Hero's Handbook. arXiv:1612.07308. Proceedings of the International School of Physics "Enrico Fermi": Course 197, Foundations of Quantum Theory. 

\item Gao, S. (ed.) (2015). Protective Measurement and Quantum Reality: Toward a New Understanding of Quantum Mechanics. Cambridge: Cambridge University Press.

\item Gao, S. (2017). The Meaning of the Wave Function: In Search of the Ontology of Quantum Mechanics. Cambridge: Cambridge University Press.

\item Gao, S. (2020). Protective Measurements and the Reality of the Wave Function. The British Journal for the Philosophy of Science, axaa004, https://doi.org/10.1093/bjps/axaa004. 

\bibitem{} Hardy, L. (2013). Are quantum states real? International Journal of Modern Physics B 27, 1345012.

\item Healey, R. (2017). Quantum-Bayesian and pragmatist views of quantum theory. In Zalta, E. N., editor, The Stanford Encyclopedia of Philosophy. Metaphysics Research
Lab, Stanford University, spring 2017 edition. 

\item  McQueen, K. J. (2017). Is QBism the future of quantum physics? arXiv:1707.02030.
% \item  Mohrho, U. (2014). QBism: a critical appraisal. arXiv:1409.3312.

\item  Mermin, N. D. (2014). QBism Puts the Scientist Back into Science. Nature 507, 421-423. 

\item  Norsen, T. (2016). Quantum solipsism and non-locality. In Bell, M. and Gao, S. (eds.), Quantum Nonlocality and Reality: 50 Years of Bell's Theorem. Cambridge University Press. pp. 204-237.

\item Piacentini, F. et al. (2017). Determining the quantum expectation value by measuring a single photon. Nature Physics 13, 1191. 

 \item Pusey, M., Barrett, J. and Rudolph, T. (2012). On the reality of the quantum state. Nature Physics 8, 475-478. 

 \item  Timpson, C. G. (2008). Quantum Bayesianism: A study. Studies in History and Philosophy of Modern Physics, 39, 579-609.

 \item  Timpson, C. G. (2021). QBism, Ontology, and Explanation. Talk given at the Mini-Workshop on QBism and the Interpretation of Quantum Theory (May 25, 2021). 
 
\item{} Vaidman, L. (2009). Protective measurements. In D. Greenberger, K. Hentschel, and F. Weinert (eds.), Compendium of Quantum Physics: Concepts, Experiments, History and Philosophy. Berlin: Springer-Verlag. pp.505-507.

\end{enumerate}
% \end{spacing}{2.0}
\end{document}